\documentclass[amsmath,amssymb,showpacs,twocolumn]{revtex4}
\usepackage{color}
\usepackage{bm}
\usepackage{graphicx}
\usepackage{epsfig}
\usepackage{bezier}
\usepackage{color}

\newcommand{\beq}{\begin{equation}}
\newcommand{\eeq}{\end{equation}}
\newcommand{\beqa}{\begin{eqnarray}}
\newcommand{\eeqa}{\end{eqnarray}}
\newcommand{\bec}{\begin{center}}
\newcommand{\eec}{\end{center}}

\def\beq{\begin{equation}}
\def\eeq{\end{equation}}
\def \be{\begin{equation}}
\def \ee{\end{equation}}
\def \bea{\begin{eqnarray}}
\def \eea{\end{eqnarray}}
\def\half{\mbox{$1\over2$}}

\def \bS{{\bf S}}
\def \bi{{\bf i}}
\def \cP{{\cal P}}

\def \cM{{\cal M}}

\begin{document}
\title{
\noindent\hfill\hbox to 1.5in{\rm  } \vskip 1pt \noindent\hfill\hbox
to 1.5in{\rm SLAC-PUB-14373 \hfill  } \vskip 1pt
\vskip 10pt
Reducing Memory Cost of Exact Diagonalization using Singular Value Decomposition}
\thanks{This work was supported by the U.~S.~DOE, Contract No.~DE-AC02-76SF00515.}
\author{Marvin Weinstein$^{ 1 }$}
\author{Assa Auerbach$^{2,3}$}
\author{V. Ravi Chandra$^3$}
\affiliation{$^{1}$SLAC National Accelerator Laboratory, Stanford, 2575 Sand Hill Road, CA 94025, USA.}
\affiliation{$^2$Department of Physics, Stanford University, Stanford CA 94306, USA.}
\affiliation{$^3$Physics Department,  Technion, Haifa, 32000, Israel}
\date{\today}
\begin{abstract}

We present a modified  Lanczos algorithm  to diagonalize lattice Hamiltonians with dramatically reduced memory requirements, {\em without restricting to variational ansatzes}.
The lattice of size $N$ is partitioned  into two subclusters. At each iteration  the Lanczos vector is projected into two sets of $n_{{\rm svd}}$
smaller subcluster vectors using  singular value decomposition. For low entanglement entropy $S_{ee}$,  (satisfied by
short range Hamiltonians), the truncation error is expected to vanish as $\exp(-n_{{\rm svd}}^{1/S_{ee}})$.
Convergence is tested for the Heisenberg model on
Kagom\'e clusters of 24, 30 and 36 sites, with no lattice symmetries exploited, using less than 15GB of dynamical memory. Generalization of the Lanczos-SVD algorithm to multiple partitioning is discussed, and comparisons to other techniques are given.
\end{abstract}
\maketitle
\section{Introduction}
Numerical  ("exact") diagonalizations  (ED) of  quantum many-body Hamiltonians on finite clusters are often used to advance our
understanding of larger lattices.  For example,
Contractor Renormalization~\cite{CORE-W,CORE-A,CORE-S} uses ED to compute the short range interactions of the effective hamiltonian.
ED on various size clusters~\cite{LauchliED} are indispensable as unbiased tests of  mean field theories and  variational wavefunctions.
They are also used to obtain short wavelength dynamical correlations~\cite{SXX}  and Chern numbers of Hall conductivity \cite{Chern}.

ED  commonly use Lanczos algorithms \cite{Lanczos-original, cullum}, to efficiently converge to the low eigenstates.
However,  for a lattice of size $N$, with $m$ states per site,  the dimension of the Lanczos vectors  (which are stored in the dynamical memory)  increases as $m^N$. Therefore, ED on larger  lattices are prevented primarily by  memory limitations, rather than processor speed.

The central idea of this paper is to significantly reduce the memory cost, in order to enable ED of  larger lattice sizes.
We  use singular value decomposition (SVD) to compress all Lanczos vectors into
sets of $2n_{\rm{svd}}$  vectors of size $m^{N/2}$.

As long as   {\em entanglement entropy} of the target eigenstates obeys
 $S_{ee}<<N/2\log(m)$~\cite{entanglementRMP}, one can greatly economize on memory  while maintaining high numerical accuracy.
 Many of the important many body Hamiltonians of condensed matter (e.g. Hubbard and Heisenberg models) have short range interactions. As a consequence, their ground states possess low entanglement entropy~\cite{EE1,EE2,EE3,EE4}. 
 
 The idea of exploiting low entanglement entropy to compress wavefunctions by  SVD, is not new. 
This is the key to the remarkable
success of density matrix renormalization group (DMRG)~\cite{whitedmrg} which has been used extensively to obtain low energy state correlations of a large variety of Hamiltonians.
DMRG is equivalent to variational minimization in the space of matrix product states~\cite{Ostlund,EE1,PPB}.
Extensions to wavefunctions with longer range correlations were given by multiscale entanglement renormalization~\cite{mera-prl}.  
Nevertheless,  sequential minimization may sometimes get ''stuck'' in  false minima and not converge to the ground state. Therefore Lanczos methods are often called for to independently test 
the variational results.

\begin{figure}[t]
\includegraphics[width=9cm,angle=0]{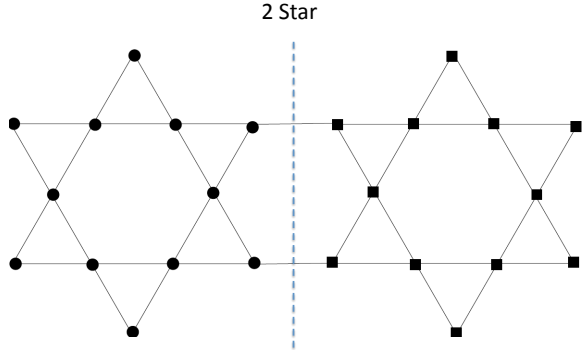}
\caption{Partitioning 24 sites Kagom\'e cluster  for application of the Lanczos-SVD algorithm. The ground state entanglement entropy is $S_{ee}=1.51$.}
\label{2star}
\end{figure}

\begin{figure}[t]
\includegraphics[width=9cm,angle=0]{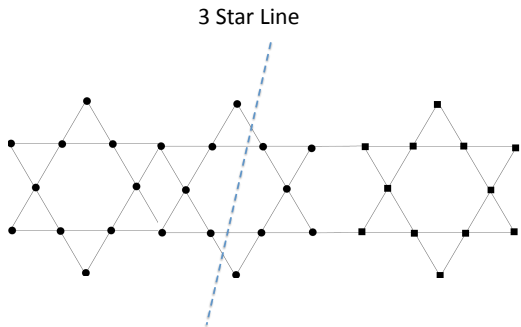}
\caption{Partitioning 36 sites for three stars in a line.  Ground state entanglement entropy is $S_{ee}\approx 1.12$.}
\label{3starline}
\end{figure}

\begin{figure}[t]
\includegraphics[width=9cm,angle=0]{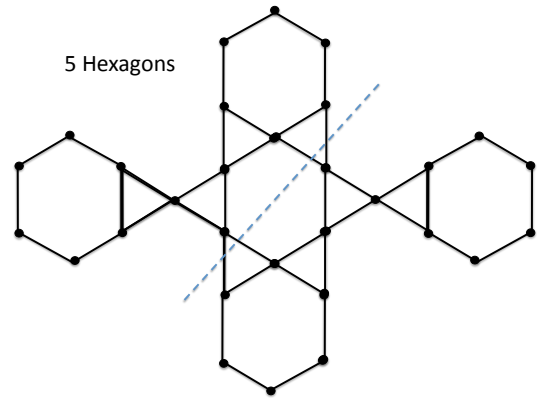}
\caption{Partitioning of 30 sites containing 5 hexagons. Ground state entanglement entropy is $S_{ee}\approx 1.27$.}
\label{5hex}
\end{figure}

\begin{figure}[t]
\includegraphics[width=9cm,angle=0]{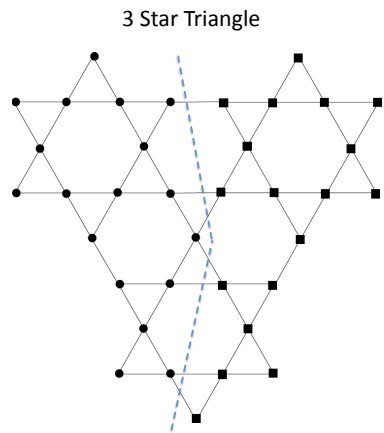}
\caption{Partitioning 36 sites. Ground state entanglement entropy is $S_{ee}\approx  2.59$.}
\label{3startriangle}
\end{figure}

The paper is organized as follows. We begin by defining the SVD for a bipartite split of the lattice, and proceed to explain in practice how perform a single Lanczos step followed by an SVD projection
which  prevents the expansion of the memory cost. We describe the intermediate matrix manipulations needed for
orthonormalizations and diagonalizations.
 In Section \ref{TE}, we estimate
the SVD truncation error after projection, as a function of $n_{{\rm svd}}$. We relate the error estimation to the bipartite entanglement entropy   $S_{ee}$, using a generic asymptotic form for the
entanglement spectrum which is based on a classical gas model. We test in detail, the convergence of the Lanczos-SVD algorithm for  the spin half  Heisenberg antiferromagnet on Kagom\'e  clusters of up to  36 sites.  The entanglement spectrum asymptotics are verified  for a partitioning of a 30 site cluster, Fig.~\ref{5hex}.
The ground state  energy of 36 sites with Lanczos-SVD converges to  relative errors of $6\times10^{-8}$ for the three star line, Fig.~\ref{3starline}, and $1.3\times10^{-4}$ for the three star triangle, Fig.~\ref{3startriangle}.
Here we use  a desktop computer with less than 15GB of memory and no lattice symmetries. The results agree with  our analytical estimate of the truncation  error. In Section \ref{sec:MP} we discuss a possible extension  of this approach to  multi-partitioning,  and estimate the optimal reduction in memory cost that could be achieved. We conclude by a discussion which elaborates on the relative advantages and disadvantages 
of  Lanczos-SVD, standard Lanczos and DMRG.

\section{ The Lanczos-SVD Step}

\subsection{Bipartite Singular Value Decomposition}
Any state $|\psi\rangle$  of the full cluster (see as an example  Fig.~\ref{2star}) can be represented in a unique  SVD form
as
\bea
|\psi \rangle &=&  \sum_{\alpha} \lambda_\alpha |\alpha\rangle_1 |\alpha\rangle_2 ,\nonumber\\
\sum_\alpha \lambda_\alpha^2&=&1,~~~~\langle \alpha|\alpha'\rangle_i = \delta_{\alpha\alpha'},
\label{psi}
\eea
where the  $\lambda_\alpha$ are positive,  and $|\alpha\rangle_i$ are "small" basis vectors of  subclusters $i=1,2$ (the subclusters on the two sides of the partitioning).
Truncating the sum into the largest $n_{{\rm svd}}$ terms defines the  SVD projector,
\be
P_{{\rm svd}}|\psi\rangle =  \sum_{\alpha=1}^{n_{{\rm svd}}} \lambda_\alpha |\alpha\rangle_1 |\alpha\rangle_2,
\label{projection-and-error}
\ee
which introduces a wavefunction error $\epsilon=\sum_{\alpha>n_{{\rm svd}} } \lambda_\alpha^2$.

\subsection{Application of $P_{{\rm svd}}H$}
 Lanczos-SVD economizes on the storage space by applying an SVD projection after each application of
 the Hamiltonian on the Lanczos vector,
 \be
|\psi\rangle'=P_{{\rm svd}} H  |\psi\rangle .
\label{psvd}
\ee
The projection entails the following computational steps.
$H$ can be written as a sum of products of the two subcluster operators,
\be
H = H^{0}_1\otimes I_2 + I_1\otimes H^{0}_2 + \sum_{\mu=3}^M H_1^\mu \otimes H_2^\mu .
\label{hameq}
\ee
where $H^{0}_i$ includes all internal interactions of subcluster $i$. $H_1^\mu \otimes H_2^\mu $ is a product of operators residing on both subclusters.
For example,  a nearest neighbor Heisenberg model ($\sum_{ij} \bS_i\cdot\bS_j$) with $K$ bonds connecting the two subclusters has  $M=2+3 K$ terms. For example in Fig.~\ref{3startriangle}, $K=7$ and $M=23$.

Acting with $H$  on $| \psi \rangle$ produces a new  state,
\be
{ H }|\psi\rangle  =  \sum_{\nu=1}^{n_{\rm{svd}} M}   |\nu)_1 |\nu )_2  ,
\label{nunu}
\ee
where the new (non orthonormal) small vectors are labeled by  the fused index $\nu=(\mu,\alpha)~$, i.e. $|\nu)_i = H^\mu_i |\alpha\rangle_i$.
The state (\ref{nunu}) lies outside the $n_{{\rm svd}}$ subspace, and we need to project it back using $P_{{\rm svd}}$ in order not to further
expand the memory cost. We first orthonormalize $|\nu)_i$ by diagonalizing the Hermitian overlap matrices (of row dimensions $n_{\rm{svd}}M$) 
\bea
 \langle  \nu' | \nu \rangle_i &=&   \left(V_i^\dagger  D_i V_i\right)_{\nu\nu'} ~~~~i=1,2   \nonumber\\
| \beta\rangle_i &=&  \sum_\nu   \left(D_i^{-\half} V_i\right)_{\beta\nu} | \nu)_i ,
\label{overlap}
\eea
where $D_i$ are diagonal and positive semidefinite, and  $V_i$ are unitary. $| \beta\rangle_i, i=1,2$ are orthonormal sets in their respective spaces. 
Thus, the new vector is given by
\bea
H|\psi\rangle&=&  \sum_{\beta \beta'}^{n_{\beta} n_{\beta'}} C_{{\beta}{\beta'}} | \beta\rangle_1 |\beta'\rangle_2 \nonumber\\
C  &=& \sqrt{D_1}V_1^* V_2^\dagger \sqrt{D_{2}}  .
\eea
Now we perform an SVD on the matrix $C$,
\be
C = \nu_c^2 U_1^t  \Lambda' U_2 ,
\label{svd-C}
\ee
where $\nu_c$ is the normalization.
$U_1^t,  U_2^\dagger$ are unitary matrices which diagonalize
the Hermitian products $CC^\dagger$ and $C^\dagger C$ respectively. After computing $\Lambda,U_1,U_2$ we
obtain the SVD form of the new state.  $\Lambda$ is diagonal and normalized to $\mbox{Tr}\Lambda^2 =1$ with positive eigenvalues $\lambda_\alpha'$. We keep only the $n_{{\rm svd}}$  largest $\lambda_\alpha'$ and obtain
\be
P_{{\rm svd}}H|\psi\rangle  =  \sum_{\alpha=1}^{n_{\rm{svd}}}  \lambda'_{\alpha}  | \alpha \rangle_1'  |\alpha \rangle_2'  ,
\label{psi'-svd}
\ee
where the new small vectors of $i=1,2$ are,
\be
|\alpha \rangle_i' =   \sum_{\nu=1}^{n_{\rm{svd}}M}\left( U_i D^{-{1\over 2}} V_i \right)_{\alpha\nu} |\nu)_i .
\ee

 \section{Entanglement spectrum and SVD truncation error} 
 \label{TE}
 The Lanczos algorithm  rotates a set of basis states  $|\psi_n\rangle$ into the lowest energy eigenstates with which  the basis  has a finite overlap.
If we choose $\epsilon$ to be much smaller than the lowest
relative energy gap,  the Lanczos-SVD  vectors converge  to a states which are  within $\epsilon$  distance from the
SVD projection of the  corresponding exact eigenstate.

To get an idea of how  $\epsilon(n_{{\rm svd}})$ converges, we must know
the "entanglement spectrum" $\{s_\alpha\}$ defined by  $ \lambda^2_\alpha \equiv e^{-s_\alpha}$.
$s_\alpha$ are  pseudo-energies of the entanglement spectrum.
A generic density of states can be modeled by a power law form
\be
\rho_p(s) =\sum_{\alpha} \delta(s-s_\alpha) = {s^p\over \Gamma(p+1)},~~~p> -1,
\label{rho_p}
\ee
which describes  the many-body density of states of
a {\em classical} gas with constant (Dulong-Petit) specific heat \cite{comment}.
$p$ counts  with the number of entangled degrees of freedom.
The corresponding entanglement entropy is easy to evaluate,
 \be
 S_{ee} = -\sum_\alpha \lambda^2 \log(\lambda^2) =\int_0^\infty ds  s \rho_p(s) e^{-s}   = p+1.
\ee
Choosing a high cut-off exponent $s_c$ such that 
\be
n_{{\rm svd}} =  \int_0^{s_c} ds \rho_p(s) \sim {s_c^{p+1} \over \Gamma(p+1) (p+1)} ,
\label{nsvd-s}
\ee
 we arrive at the error estimate
\be
\epsilon = \int_{s_c}^\infty ds \rho_p(s) e^{-s}  \simeq  {s_c^p e^{-s_c}\over \Gamma(p+1)}
\label{epsilon-s}
\ee
Combining Eqs.~(\ref{nsvd-s}) and (\ref{epsilon-s}), yields for $S_{ee} >>  1$, the asymptotic expression
\be
\epsilon   \sim   n_{svd} S_{ee}  e^{- S_{ee}(n_{{\rm svd}})^{1\over S_{ee}} }
\label{err-est}
\ee
Hence by  choosing the ratio $n_{{\rm svd}} /  e^{S_{ee}} >>1$ one ensures an exponentially  small truncation error.

\subsection*{Numerical entanglement spectrum}

\begin{figure}[t]
\includegraphics[width=9cm,angle=0]{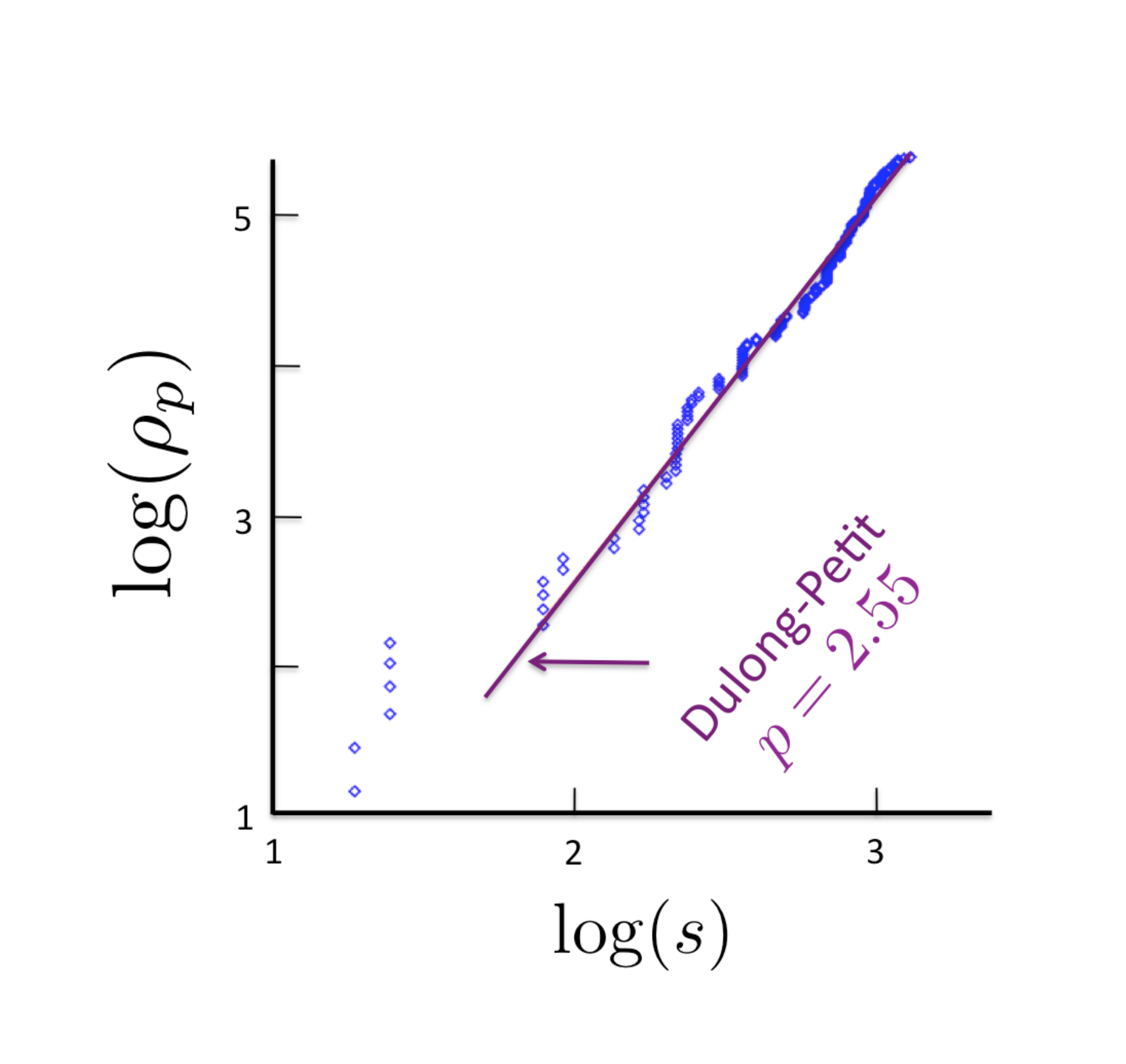}
\caption{Entanglement spectrum. We extract the asymptotic behavior of the entanglement density of states $\rho_p$, for the first excited state of the five hexagon cluster of Fig. \ref{5hex}.
 The line depicts a pure power law at large pseudoenergies $s$, consistent with the Dulong-Petit form given in Eq.~(\ref{rho_p}).}
\label{30ES}
\end{figure}

In Figure \ref{30ES},  we plot the entanglement spectrum for the first excited eigenstate of the 30 site, five hexagon Kagom\'e cluster depicted in Figure \ref{5hex}.
The log-log plot demonstrates the asymptotic power law  density of states $\sim s^p$, where $p\approx 2.55$. 
We note that for this system we are in the low $S_{ee}$ regime and hence the difference between the fitted value of  $p+1$ and the entanglement entropy of $1.27$.
Nevertheless, the   density of states at high pseudo-energies  extrapolates well  to the asymptotic power law behavior of 
a classical gas as modelled in this section.

\section{Implementation of the Lanczos-SVD iteration}
The Lanczos-SVD routine proceeds as follows: We initialize $|\psi\rangle^{(0)}$ as a direct product of the two subcluster states.
We compute  $(\cP_{\rm{\rm{svd}}}H)^n|\psi^{(0)}\rangle=|\psi^{(n)})$ as described above.
Since our method is economical in memory, we can afford to retain
$L$ sequential Lanczos vectors  $|\psi^{(n)}) , |\psi^{(n+1)}) ,\ldots |\psi^{(n+L)})$, which speeds up the convergence with
iteration number considerably. (If memory is scarce, one could use the slower method of keeping only two Lanczos vectors).

 Now, we compute the overlap matrix and orthonormalize this  set of Lanczos vectors. This
produces  a "rotating basis" of dimension $L$
\be
|\varphi^{(i)}\rangle=\sum_{n'=n}^{n+L} A_{i n'} |\psi^{(n')}\rangle ,~~~~ \langle \varphi^{(i)}|\varphi^{(j)}\rangle=\delta_{ij}  ,
\ee
where $A$ are the coefficients determined by diagonalizing the overlap matrix (see e.g. Eq.~(\ref{overlap})).

Subsequently, we compute the matrix elements of the reduced Hamiltonian,
\be
H_{ij} = \langle \varphi^{(i)}|H|\varphi^{(j)} \rangle ,~~~~i,j=1,\ldots L.
\ee

The  reduced Hamiltonian matrix is diagonalized. Its lowest eigenvalue and eigenvector yield the
best approximation to the ground state at this level of iteration~\cite{weinsteincoreprd1993}.
We bring the resulting wavefunction to the SVD form, again truncated into $n_{\rm{\rm{svd}}}$ terms. It becomes the
new initial state $|\psi^{(n+L+1)}\rangle$  for the next Lanczos iteration.
Excited states can be calculated by starting with an initial state orthogonal to
the converged lower energy states.

\begin{figure}[t]
\includegraphics[width=8cm,angle=0]{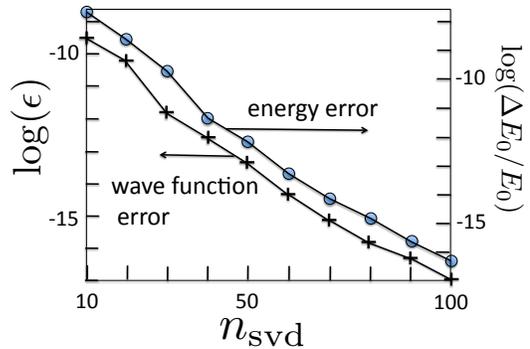}
\caption{Lanczos-SVD truncation errors, for the 24 sites Kagom\'e cluster of Fig.~\ref{2star}.
$n_{\rm{\rm{svd}}}$ is the number of retained  SVD states of the ground state (\ref{psvd}).
$\epsilon$ is the wavefunction error Eq.~(\ref{projection-and-error}) and following text). $\Delta E_0/E_0$  is the relative error in the the Lanczos-SVD ground state energy
as
compared to the exact (standard Lanczos)  result. The rapid decay of the errors  for low values of  $n_{\rm{\rm{svd}}}<< 2^{12}$  is due to the low entanglement entropy, see  Section \ref{TE}. }
\label{err-24}
\end{figure}


\begin{figure}[t]
\includegraphics[width=9cm,angle=0]{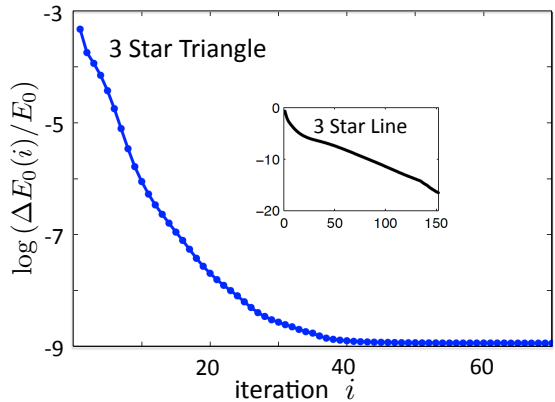}
\caption{Lanczos-SVD error convergence as a function of iteration, for  36 site Kagom\'e clusters. The main plot is for Fig.~\ref{3startriangle}, and the inset is for Fig.~\ref{3starline}.
$\Delta E_0(i)$ is difference between the  energy at iteration $i$ and the converged ED result given by standard Lanczos~\cite{sylvain36}.  For both clusters we used $n_{\rm{svd}}=200$ 
and $L=4$ Lanczos vectors.  }
\label{err-36}
\end{figure}

\section{Numerical tests}
 For the spin-half Heisenberg antiferromagnet
\be
H=\sum_{\langle ij\rangle} \bS_i \cdot \bS_j,
\ee
we  tested the convergence of the Lanczos-SVD algorithm
on Kagom\'e clusters  depicted in Figs.~\ref{2star}, \ref{3starline}, \ref{5hex} and \ref{3startriangle}.

In Fig.~\ref{err-24},  we plot the Lanczos-SVD  truncation error for the ground state wave function and  energy,
as a function of $n_{{\rm svd}}$.   The errors decrease rapidly on the logarithmic scale 
as expected by Eq. (\ref{error-est}), arriving at  $\approx 10^{-13}$ for $n_{{\rm svd}}=200$. 

For the 30 site Kagom\'e cluster  depicted in  Fig~\ref{30sites}, the energies of the four lowest S=0 eigenstates, and  the first triplet S=1 state,  converged to 
a very high  accuracy of $10^{-11}$ using  $n_{{\rm svd}}=200$.

 In Fig.~ \ref{err-36} we show the convergence of  ground state energy of Lanczos-SVD  versus iteration for clusters of 36 sites. We use
$n_{\rm{\rm{svd}}}=200$, and $L=4$.  For the   three star triangle, the exact ground state energy as
determined by standard Lanczos is $E_{0} (36 {\rm{sites}}) = -14.859397$  \cite{sylvain36}. The entanglement entropy 
is $S_{ee}\approx 2.5$. The calculation converges to relative energy accuracy of $1.3\times 10^{-4}$,  
and an SVD truncation error of similar magnitude.

In the inset of Fig. \ref{err-36} we show a much smaller error  for the three star line of Fig.~\ref{3starline}, which converged to a relative error of $ 6.3\times 10^{-8}$.    
This is to be expected since the  entanglement entropy of the
linear arrangement of the three stars is only $S_{ee}\approx 1.12$.

The numerical tests were therefore consistent with Eq ~(\ref{err-est}).

All the above calculations were performed using multi-core workstations.
The maximum memory usage was kept under 15 GB of memory, even though no lattice 
symmetries were implemented in the computations. The time required for the most intensive
calculations (36 sites, $n_{{\rm svd}}=200$) on using parallelisation with 16 cores was a little more 
than 70 minutes per iteration Fig \ref{3startriangle} and about 35 minutess for Fig. \ref{3starline}. 
A serial MAPLE 15 implementation used in the computations for the 30 site case
took about one day for each eigenstate for the same $n_{{\rm svd}}$.

\section{Extension to multiple partitioning of large lattices.}
\label{sec:MP}
The Lanczos-SVD  compresses the memory requirement by a single division of  the cluster into two subclusters $i=1,2$.
This idea could be extended to recursive partitioning~\cite{mera-prl}. For the sake of crude memory estimation,  each small vector (e.g. $|\alpha\rangle_i$ in Eq.~(\ref{psi}) can be decomposed into $n_{{\rm svd}}$  products  of  even smaller subcluster vectors.
If the SVD is thus  iterated $p$ times, one obtains a representation in terms of small vectors of $P=2^p$  subclusters.  $|\psi\rangle$
is thus stored in terms of a set of the smallest vectors.  The memory cost after applying a Lanczos step is as follows.

\begin{figure}[t]
\includegraphics[width=8cm,angle=0]{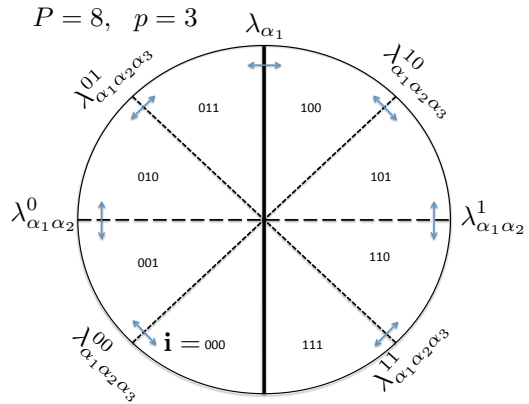}
\caption{Multiple subclusters}
\label{pie}
\end{figure}

For concreteness, let us consider a two dimensional disk of radius $R>>1$, containing
$N\simeq \pi R^2$ sites of spin half,  divided into $P$ equal sections as shown in Fig.\ref{pie}. The sections are labeled by a
 binary number ${\bi}=(i_1,i_2,\ldots i_p)$,  $i_k=0,1$.
The recursive SVD decomposition yields the expression
\be
|\psi\rangle = \sum_{\alpha_1,\alpha_2\ldots \alpha_p} \lambda_{\alpha_1} \lambda^{i_1}_{\alpha_1,\alpha_2}\cdot\cdot   \lambda^{i_1,\ldots i_{p-1}}_{\alpha_1,\ldots\alpha_p} \prod_{{\bi}} |\alpha_1,\ldots \alpha_{p}\rangle_{{\bi}}  .
\ee
The SVD weights $\lambda^{{\bi}}$ are labeled according to the boundaries they describe, as shown in Fig.~(\ref{pie}).
Each $\alpha_i$ runs over $n_{{\rm{svd}}}$ numbers, which means that each section $\bi$ is represented by $n_{{\rm{svd}}}^p$  vectors of dimension $2^{N/P}$.
By the "area law" $S_{ee}\propto R$ on each boundary.  As shown before,  we must retain 
$n_{{\rm svd}} \sim e^{cR}$ terms in each SVD, where $c(\epsilon)>1 $ in order to achieve a desired truncation error $\epsilon$.

After applying $H$ to $|\psi\rangle$, we generate a factor of
$M\approx 6R$ more small vectors.  Thus we should store   $6PR  n_{\rm{svd}}^{p}$ small vectors.
Thus the memory cost is  
\be
 \cM_c  \approx   6PR \exp\left( {cR  \log(P)}  + {\pi R^2\log(2) \over   P}\right)  .
\ee
Minimizing $\cM_c(P)$ one finds the optimal partitioning $P^{opt}$, and the optimal memory cost $\cM^{opt}$ at large $N$ to scale as
\be
P^{opt} \approx {\pi \log(2) R\over c},~~~\cM_c^{opt} \sim N  e^{{c\over 2 }\left(\sqrt{N/\pi}\log(N/\pi)-2\right)}  .
\ee
This  would amount to a significant  compression of memory as compared to standard Lanczos $\cM\sim 2^N$.
The remaining  challenge is to speed up the significantly larger computational time needed to orthonormalize and SVD  large  sets of small vectors.

\section{Discussion}

In this paper we have exploited the SVD to expand the lattice sizes which can be treated by the Lanczos algorithm for ED.
Relative to standard Lanczos, Lanczos-SVD  demands longer  computation time due to additional matrix manipulations.

{\em Does the SVD projection interfere with the Lanczos  convergence?}. 
The SVD projection  introduces a truncation error
in the rotating vector. However, the Lanczos vector  is rotated toward the ground state (or some other target state)  when  the 
relative energy splitting to neighboring eigenstates is  larger than the truncation error.  This rotation ceases once
the energy has converged within  the SVD error. Increasing $n_{{\rm svd}}$ will allow further convergence.
It is  simplest to think about the SVD projection in the same footing as an additional floating point  error which  limits the numerical convergence of standard Lanczos algorithms.
In computing Fig.~\ref{err-36}  we have indeed verified  that the implementation of  SVD projection in each iteration, does not slow down the energy convergence per  each  Lanczos step.

Computing time of Lanczos-SVD may  be significantly reduced by using multiple cores
and by parallelizing the code. In particular the overlap calculations, which are currently the most time-consuming, are easily parallelizeable.
In standard Lanczos, memory reduction can be achieved by exploiting lattice and spin symmetries, which demands special boundary conditions and extensive programming. 
Lanczos-SVD  is therefore simpler to implement, and can address arbitrary boundary conditions.

{\em It is often asked what advantage a Lanczos based ED has over DMRG and related variational methods?}  The answer of course depends on the purpose of the calculation.
DMRG has proven very efficient (especially for ground state correlations) for  larger lattices than Lanczos methods can address. However, in practical applications, DMRG
advances  toward the ground state by
{\em sequential} minimizations, e.g. "sweeping"
the parameters of the wave function in real space.    In cases of high frustration and competing phases (e.g. near a phase transition),   sweeping methods
can get stuck  in  metastable states~\cite{DMRG2D}. (Consider e.g. the difficulty of getting rid of defects in a phase separated system by sweeping methods).   

The Lanczos step, on the other hand,  does not necessarily move the state in the direction of 
maximal slope. Instead, if numerical accuracy is sufficient,  it steadily rotates the Lanczos vector  toward the true ground state (or some other target eigenstate). 

Lanczos-SVD can therefore be used to determine the ground state and low excitations of limited size  clusters  with well controlled accuracy. 
It could be used to check DMRG convergence and test variational ansatzes.
As mentioned in the introduction, a primary purpose for using ED on small clusters is to derive an effective Hamiltonian  
by the CORE method\cite{CORE-future}. The CORE effective Hamiltonian can then be studied on the coarse grained lattice by iterating CORE, or by variational methods. 

\subsection*{ Acknowledgments} We thank Dan Arovas, Yosi Avron, Sylvain Capponi,  Steve Kivelson, Israel Klich,  Netanel Lindner and Steve White for insightful conversations.
AA acknowledges support from Israel Science Foundation, and US-Israel Binational Science Foundation, and the hospitalilty of Aspen Center For Physics.

\end{document}